# Collective treadmilling in fire ant rafts permits sustained protrusion growth

Robert J. Wagner[1], Kristen Such[2], Ethan Hobbs[3], Franck J. Vernerey[4]*

[1]University of Colorado, Boulder. Mechanical Engineering Department. Material Science and Engineering Program.

[2]University of Colorado, Boulder. Mechanical Engineer Department.

[3]University of Colorado, Boulder. Computer Science Department. Interdisciplinary Quantitative Biology Program.

[4]University of Colorado, Boulder. Mechanical Engineer Department. Material Science and Engineering Program. Email: Franck.Vernerey@Colorado.edu

*Franck J. Vernerey

**Email:**  Franck.Vernerey@Colorado.edu

**Abstract:** Condensed active matter is exemplary for its capacity to morph and exhibit internal flows despite remaining cohered. (1) To facilitate understanding of this ability, we investigate the cause of finger-like protrusions that emerge from super-organismal, aggregated rafts of fire ants (*Solenopsis invicta*). While these features are easily observed, what permits their recuring initiation, growth, and recession is not immediately clear. Ants rafts are comprised of a floating, structural network of interconnected ants on which a layer of freely active ants walks. We show here that sustained shape evolution is permitted by treadmilling defined by the competition between perpetual raft contraction due to displacement of bulk structural ants into the active layer, and outwards raft expansion due to deposition of free ants into the structural network at the edges. Furthermore, we see that protrusions emerge due to asymmetries in the edge deposition rate of surface ants, and we provide strong evidence that these asymmetries occur stochastically due to wall accumulation effects and local alignment interactions. Together these effects permit the cooperative, yet spontaneous formation of protrusions that fire ants utilize for functional exploration and to escape flooded environments. Ant raft dynamics mirror the treadmilling that facilitates morphogenesis and motility of cytoskeletons (2, 3), thus providing another example in which reshaping of condensed active matter is explained by free constituent transport, yet whose continuance relies on perpetual phase transition between structural and free members.

**Significance Statement:** Cooperative behavior in populations of active particles often drives complex emergent functions. In this context, rafts created by *Solenopsis invicta* (fire ants) during floods exhibit a peculiar behavior. While the structural networks of these rafts were previously believed static, we show here that they undergo contractile flow that fuels a flux of ants into a dispersed active layer on their surfaces. These active ants then rebind to the network's structural edge permitting growth of tether-like protrusions that colonies can use as escape routes. Our results provide a remarkable example of how an entangled active system achieves shape change and spontaneously explores its environment through cooperative behavior, thereby providing insight for the innovation of active materials and swarm robotics.

**Main Text**

**Introduction**

Collective emergent behavior is a remarkable and omnipresent feature of living systems that often results in functions like motility of aggregations, self-healing of tissues, and morphing of swarms (1, 4). Nature is brimming with examples of collectively interacting organisms that exists in both dispersed, fluid-like phases (such as herds of animals evading predators) and condensed, solid-like phases (such as clusters of eukaryotes exploring their environment) (5, 6). While the local interactions of dispersed superorganisms have been well-studied and distilled down to very simple sets of rules such as those of the Vicsek model (7), much remains unknown regarding the flow of elementary constituents in their condensed counterparts or how these superorganisms remain cohered while still collectively undergoing substantial changes in shape.

One common category of such organisms, favored for their macroscopic size and ease of observation, is aggregations of insects, including those of the red imported fire ant (*Solenopsis invicta*). Fire ants condense into buoyant aggregations comprised of worker ant bodies when their habitats are flooded and form complex 3D structures such as towers (8, 9) or, as we observe here, 2D land-bridges (**Fig. 1.A**). While the behavior and flow of 3D ant towers has been well-studied (10, 11), the formation of fingering protrusions has, to our knowledge, neither been documented nor explained in existing literature. In this work, we experimentally examine fire ant rafts to characterize the growth and shape-evolution of these formations and uncover the dynamical treadmilling of ant rafts that allows these protrusions to emerge and decay cyclically and sustainedly over the span of hours (**Fig. 1.B-C**). In doing so, we identify ant rafts as a biphasic system whose shape evolves due to not only transport of unbound surface ants, but also global contraction of their structural network. This demonstrates that convection is a feature common to both free and entangled active matter systems. Furthermore, we identify the underlying mechanisms that drive tether-like protrusion growth.

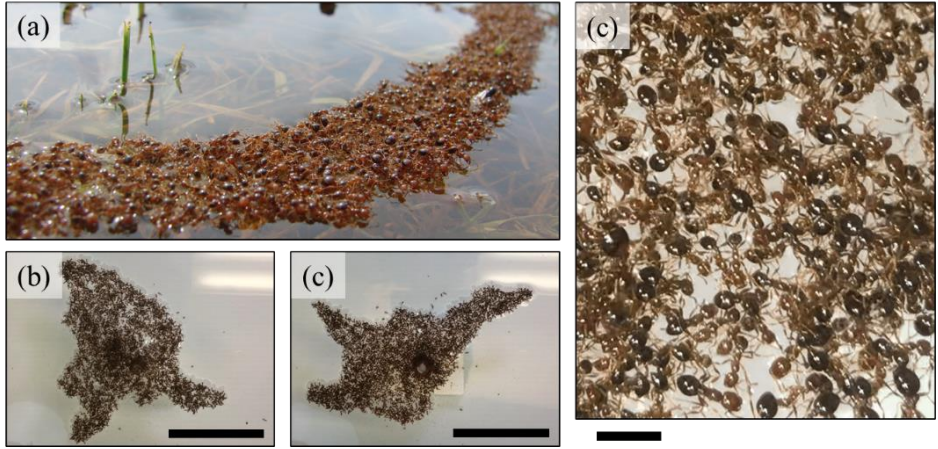

**Figure 1. Images of networked fire ant rafts display dynamic structures: (A)** Fire ants in nature form rafts that sometimes create long, tether-like protrusions, as shown, which serve as land bridges. A top view of an experimental ant raft is depicted at the **(B)** start and **(C)** end of a roughly one-hour duration to illustrate that the global shape can change significantly despite roughly conserved raft area, and that protrusions emerge and are reclaimed over hour timescales. Scale bars represent $20\ \ell$ where $\ell \equiv 1\ average\ ant\ body\ length$. **(D)** Despite their significant dynamics, ant rafts remain cohered as a networked material crosslinked by ant-to-ant bonds. The scale bar represents $1\ \ell$.

## Results

**Ant rafts as biphasic active matter:** To begin our experiments, we collected 3 to 10 g of worker ants and transported them to a container of water in which an acrylic rod stemming vertically from the base of the container anchored the rafts to a still reference frame. The ants enveloped and nucleated to the acrylic rod at which point an ant raft was permitted to reach a pseudo-steady state and filmed over second and hour timescales. Confirming the observations of both Mlot, *et al.* (9) and Adams, *et al.* (8) we see that fire ant rafts are comprised of a bounded network of ants cohered via tarsi-to-tarsi bonds (8, 12) (**Fig. 1.D**) that floats on the water. On top of this network, a layer of unbound ants walks freely (see **Movie S1** for visual illustration). The packing fraction or ant density of the raft network was found to be roughly constant at $\rho_r = 0.304 \pm 0.018$ ants mm$^{-2}$. While the density of the unbound ants was difficult to measure due to heterogeneity and clustering, surface density $\rho_s$ was measured in regions with sufficient visual contrast. We found that $\rho_s = 0.072 \pm 0.006$ ants mm$^{-2}$, indicating an estimated surface packing fraction of $\phi \approx 0.240$ unbound ants per structural ants. Although the actual surface packing fraction may be higher than this value, a significant majority of fire ants in these experiments acted as structural members of its floating base. In the remainder of this work, we will refer to ant rafts as biphasic active matter given their comprisal of both a monolayered structural network phase, as well as a vapor-like surface phase.

**Rafts expand due to edge deposition:** Upon being placed into the water as roughly 3D spheroids, ant rafts will spread out rapidly. This feature was studied by Mlot, *et al.* (13) in which they modeled surface ants as particles undergoing Brownian diffusion. Mlot, *et al.* (9) reported that unbound ants walk on the surface of the structural raft until they encounter the edge at which point they either bank off said edge or choose to deposit into the structural layer with some constant probability. Assuming a roughly conserved structural network density, we measured the global, areal raft expansion and estimated that unbound ants park at a normalized rate of $\alpha \approx 0.02$ binding events per minute per structural ant (or $\gamma \approx 0.29$ binding events per minute per unit body length of raft perimeter). We measured the expansion area, shaded cyan in **Fig. 1.A**, by tracing ants that were originally in the structural network's perimeter, outlined red in **Fig. 1.A**, and calculating the additionally deposited area to the rafts' leading edges. Local observations indeed confirm that this outwards expansion was due to deposition of surface ants into the structural layer at the edges of the raft (see **Movie S2** for example). If the raft expanded unchecked this would correspond to areal raft expansion on the order of 2% min$^{-1}$ until the number of free ants was depleted, and a static raft area was reached. As such, this mechanism alone can explain neither the breaks in local edge symmetry that lead to tether-like growths, nor the dynamical pseudo-steady state observed here in which these growths repeatedly emerge from and are reclaimed by the raft over the span of hours. To better understand the full scope of what drives these features, we begin by quantifying the transport of surface ants in greater detail.

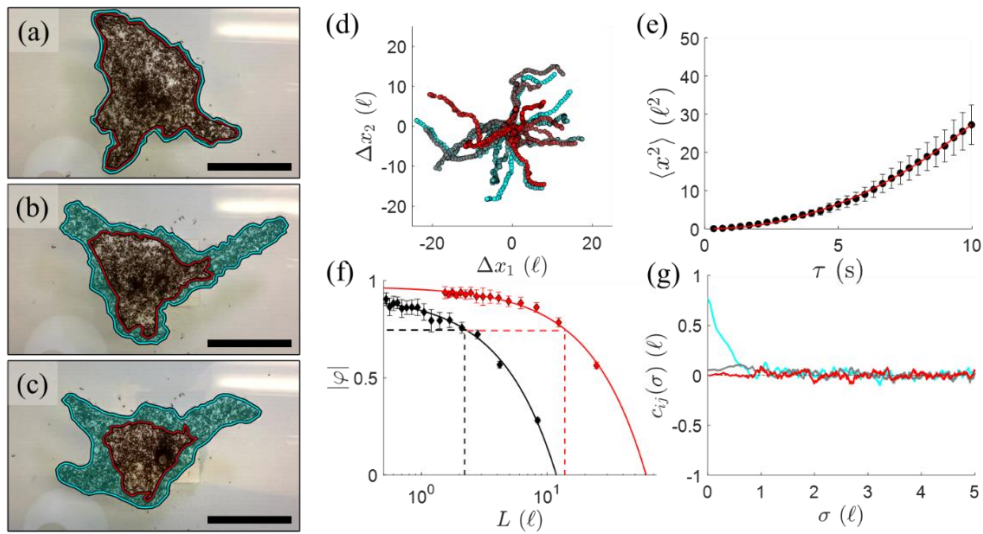

**Figure 2. Trajectory analysis of non-edge-encountering surface ants: (A-C)** An ant raft is depicted at the **(A)** start, **(B)** middle, and **(C)** end of a 54-minute duration. The red outline represents ants that were originally in the edge of the raft at the start of the timespan and the region shaded in cyan represents the newly deposited raft area. Scale bars represent 20 $\ell$. **(D)** The end-to-end trajectory of 38 freely active ants image tracked in a domain of width $\sim 50\ \ell$ over a duration of roughly 30 seconds visually indicates that they appear to move isotropically with no clearly biased direction. **(E)** Mean square displacement $\langle x^2 \rangle$ plotted with respect to the time interval of measurement $\tau$ illustrates a divergence in slope indicating super-diffusive behavior. The red curve represents the least-squares regression fit of the form $\langle x^2 \rangle = 4D\tau^\xi$ (see **Fig S2** for full data sets). Error bars represent standard error. **(F)** $|\varphi|$ is plotted with respect to the rectilinear domain size $L$ in which it was measured for two separate experimental sets of ants (black and red). A strong degree of order in the velocity of surface ants ($|\varphi| > 0.75$) occurred when $L < 2.2\ \ell$ (black dotted lines) and $L < 13.8\ \ell$ (red dotted lines), for the two experiments, respectively. Error bars represent standard error. **(G)** The moving average of $c_{ij}$ is plotted with respect to separation distance for $\tau = 0\ s$ (cyan), $\tau = 1\ s$ (grey), and $\tau = 10\ s$ (red). The moving average window was set to 1 $\ell$ to reduce noise for transparency.

Here we confirm that surface ants on the bulk of the raft (*i.e.,* that do not encounter the raft edge) appear to walk isotropically (**Fig. 2.A**) with a persistence length $l_p$ on the order of 20 $\ell$, indeed indicating that locomoting ants will regularly encounter the edge of rafts whose confining dimensions are comparable to this length scale (14). The experimental rafts used herein have confining dimensions on the order of 20 to 50 $\ell$, suggesting that there will exist a persistent flux of ants from the bulk of the raft to the edge (15). Although our estimate of $l_p$ affirms that – at least some – surface ants sustain correlated trajectories over the order of $\sim 10\ \ell$, our methods of estimating $l_p$ are extrapolatory, and so to better characterize ants' trajectories we also examine their mean-square displacements ($\langle x^2 \rangle$). We find that surface ants have an average measured diffusion coefficient in the range of $\bar{D} = 0.11\ \ell^2\ s^{-1}$ to $0.15\ \ell^2\ s^{-1}$ ($0.9 \times 10^{-6}$ to $1.2 \times 10^{-6}\ m^2\ s^{-1}$) depending on the experiment and sample, placing the order of free ants' diffusivity near that of gaseous particles. Significantly, the ants do not diffuse randomly as previously reported (9, 13). Instead, ants generally diffuse anomalously according to a power law $\langle x^2 \rangle = 4D\tau^\xi$, where the average scaling coefficient is $\bar{\xi} \approx 1.92$, suggesting that the ants are typically super-diffusive ($\xi > 1$) and in some instances, hyper-ballistic ($\xi > 2$) (16, 17) (See **Fig. 2.B** for ensemble average). Worth noting is that for a given ant trajectory (see **Fig. S2** for extended data), ants undergo instances of transition to sub-diffusive behavior ($\xi < 1$). This occasional alternation is likely due to factors such as ants' tendency to stop and clean themselves at times, and the formation of ant clusters in which ants get entangled over second to minute timescales. Regardless, the prevailing behavior appears to be that of super-diffusivity. Super-diffusivity is a common feature of biological systems (18) and self-propelled particles in general (19) as it is generally indicative of current-induced drift (20). In this case it is likely that local currents emerge over some length scale due to inter-ant interactions providing early evidence that ants exhibit some degree of collective and correlated motion.

To evaluate the degree and length scale over which order in ant trajectories exists, we examine the normalized velocity order parameter $|\varphi| = \langle v(t) \rangle_N / \langle |v(t)| \rangle_N$ of ants within successively smaller domains of square dimension $L$ (**Fig. 2.C**). For purely isotropic motion, we expect that $|\varphi| \approx 0$, while for completely synchronous motion of all particles in our domain, we expect that $|\varphi| \rightarrow 1$ (4). We see that across two different experiments, $|\varphi|$ scales linearly with respect to $L^{-1}$. However, between these experiments the length scale at which $|\varphi|$ diminishes differs greatly, with ants in one experimental dataset having strongly ordered trajectories ($|\varphi| \geq 0.75$) over a domain size of $L \leq 14$ $\ell$ while ants in the other experiment display this degree of order only when $L \leq 2$ $\ell$. Although the evolution of this order parameter will vary significantly between sample domains, likely due factors such as raft geometry and surface packing fraction (21–23), it is evident that surface ants on rafts demonstrate some degree of collective motion and that the length-scale of ordered movement can transcend that of their body length by an order of magnitude. However, $|\varphi|$ alone does not discern the nature of alignment interactions nor the precise length-scale over which they occur.

To identify the length-scale of correlation between the trajectory of ants neighboring, we examine the pair-wise directional correlation between the velocities of surface ants over a time span $\tau$, given by $c_{ij}(\sigma) = \langle \hat{v}_i(t) \cdot \hat{v}_j(t + \tau) \rangle$ (**Fig. 2.D**) where $c_{ij} \rightarrow 1$ indicates strong correlation. Here, $v_i(t)$ and $v_j(t + \tau)$, are the velocities of ant $i$ (at time $t$) and ant $j$ (at time $t + \tau$), respectively, and $\sigma$ is the ant-to-ant separation distance. The delay in this directional correlation is commonly used to identify leaders and followers in the collective motion of free self-propelled particles but here – plotting $c_{ij}$ with respect to $\sigma$ – it may be used to identify the length and time scales above which ants lose their directional alignment (4). Examining the moving average of $c_{ij}(\sigma)$ (taken over a window of 1 $\ell$ to reduce noise), there appears to be no significant correlation above a length scale of ~ 1 $\ell$, regardless of $\tau$, suggesting negligible alignment effects between ants separated by more than their own body length. It is also true that over a span of $\tau \geq 1\,s$, there appears to be no correlation in direction between ants regardless of their location. The only spatiotemporal separation for which a significant directional correlation occurs is in the regime $\sigma < 0.75\,\ell$ and $\tau < 1\,s$, which suggests synchronized movement only reliably occurs between ants that are in (or nearly in) instantaneous contact. This is strongly indicative of steric-like alignment interactions between surface ants. These local interactions may give rise to local alignment that – depending on surface packing fraction – is reflected in values of $|\varphi|$ measured on the order of ant persistence length (~ 10 $\ell$).

Later in this work, we examine how these local alignment effects contribute to the initiation and runaway growth of protrusions; however, first we reconcile how raft dynamics permit sustained shape evolution. Recalling that outwards expansion alone cannot explain this phenomenon, we turn to examining the behavior of the structural network layer. Simply reexamining **Fig. 1.A-C**, it is immediately clear that the area circumscribed by the set of ants initially in the raft perimeter (outlined in red) depreciates in time, indicating that some contractile mechanism occurs within these systems.

**Structural contraction counteracts growth:** Although ant rafts' structural networks appear to be amorphous solids at first glance, examination of time-lapsed footage reveals that these network flows visibly. Specifically, the structural network layer contracts inwards towards the fixed anchor point of our experiments in time (See **Movie S3**). To quantify this phenomenon, we began by image-tracking sets of bound ants originally at the perimeter as they flowed inwards (**Fig. 3.A-B**). The areas circumscribed by these ants decay exponentially in time (**Fig. 3.C**) according to $A = A_0 e^{-2\dot{\varepsilon}t}$ where $A_0$ represents the initial reference area and $\dot{\varepsilon}$ is the 1D strain rate assuming isotropic contraction. Applying a least-squares regression fit to this data indicates that $\dot{\varepsilon} \approx 1.75\ \%\ \text{min}^{-1}$ ($R^2 = 1.00$). To confirm the isotropy of contraction, we also plotted the radial component of ant speed with respect to distance from the acrylic anchor point for each set of ants (colored data in **Fig. 3.E**) and found that the velocity towards our anchor point was roughly linear with respect to distance. Thus, taking the slope of the linear regression of this data, we estimated that the mean radial strain rate is on the order of $\langle\dot{\varepsilon}\rangle = 1.82 \pm 0.03\ \%\ \text{min}^{-1}$ for the experiment depicted in **Fig. 3**, which is within 2 % of the value coarsely estimated through the areal decay rate, thereby confirming roughly isotropic contraction (*i.e.*, that the transverse contractile rate is roughly the same as the radial contractile rate). To substantiate the results of manual image tracking, we also conducted particle image velocimetry (PIV) (24, 25) on a continuous material region of interest within the raft (**Fig. 3.D**). Again, examining the velocity component vectored towards the still reference frame we found that the contractile speed is roughly linear with respect to distance from the anchor point and that the slope of the linear regression of this line is $\dot{\varepsilon} = 1.75 \pm 0.01\ \%\ \text{min}^{-1}$ ($R^2 = 0.97$), which closely agrees with the values from our manual method.

That the contraction is isotropic, and the speed of contractile strain is roughly linear with respect to the distance from the still reference frame, both suggests that there is, in fact, a constant strain rate occurring throughout the material and that the primary mechanism of contraction in ant rafts originates within the bulk material as opposed to, for example, the interface between the raft and the anchoring rod. Interestingly, despite contraction, network density far from the acrylic rod ($\rho_r = 0.304 \pm 0.018$ ants mm$^{-2}$) was roughly conserved throughout the experiments, mandating that there is a flux of ants out of the structural layer. Upon closer examination, we indeed observed instances of ants bound to the raft network unbinding and becoming part of the pedestrian surface layer (see **Movie S4**). Assuming roughly constant structural network density, we quantified the unbinding rate, $\delta$, through the relationship $\delta = -2\rho_r\dot{\varepsilon}$ (see **Methods and Materials** for details) to find that across experiments, ants unparked at a rate of $\delta \approx 2-3\%\ \text{min}^{-1}$, counteracting and nearly balancing the global expansion rate $\alpha \approx 2\ \%\ \text{min}^{-1}$ we reported earlier.

It is the combination of global raft expansion (due to microstructural deposition of unbound ants into the structural layer at the edges) and global raft contraction (due to microstructural dislocation of bound ants into the freely active layer in the bulk) that sustains ant rafts' ability to change their shape over long timescales. Together these phenomena culminate in global, torus-like treadmilling (illustrated schematically in **Fig. 4**) likened to that of the cytoskeletal networks of cells (2), and which depends on the biphasic nature of ant rafts. This treadmilling permits dynamical shape evolution due to recycling phase transition of both structural and surface ants, thus it is prerequisite to the sustained formation of protrusions. However, the detailed causes of protrusion initiation and unstable runaway growth remain unclear. To unveil these mechanisms, we revisit the properties of unbound surface ants.

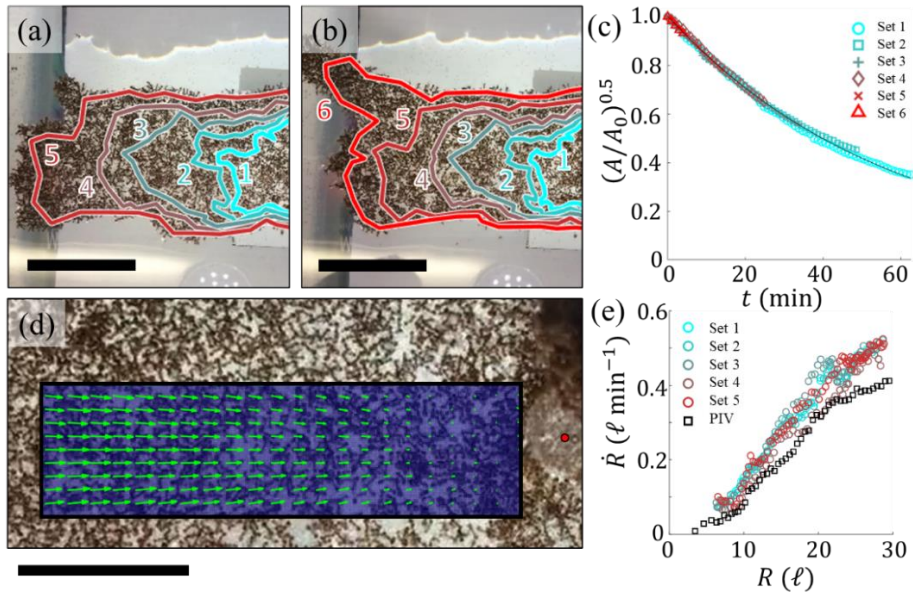

**Figure 3. Quantifying structural retraction: (A-B)** A top view of an experimental raft is illustrated at the start **(A)** and end **(B)** of a roughly 8 min duration. The perimeter is traced every 15 minutes and outlined by numbered, colored contours with 1 representing the oldest set of ants and 6 representing the newest. The scale bars represent 20 $\ell$. **(C)** The square root of the ratio of the area circumscribed by each set of ants, $A$, to initial area, $A_0$, is plotted with respect to time and used to estimate $\dot{\varepsilon}$ according to $A = A_0 e^{-2\dot{\varepsilon}t}$. **(D)** A continuous material region of interest was selected from the largest of experiments on which to conduct PIV. The resulting velocity field is shown. To eliminate noise due to raft spin, only the radial component of the velocity (*i.e.,* that vectored towards the anchor point of the raft depicted as a red dot) is shown. To reduce temporal noise the data depicted is averaged over the full PIV analysis duration (roughly 13 min). The scale bar represents 10 $\ell$. **(E)** Results of $\dot{R}$ from manual image tracking (circles in a cyan to red color gradient) and PIV (black squares) are plotted with respect to $R$. The slope of the linear regression to each data set (not shown) is taken as $\dot{\varepsilon}$. Data shown from manual tracking represents the contractile speed of every ant sampled about the entire azimuthal domain (*i.e.*, the full edge of the raft domain). Data from PIV is presented from every point measured in the region of interest from PIV.

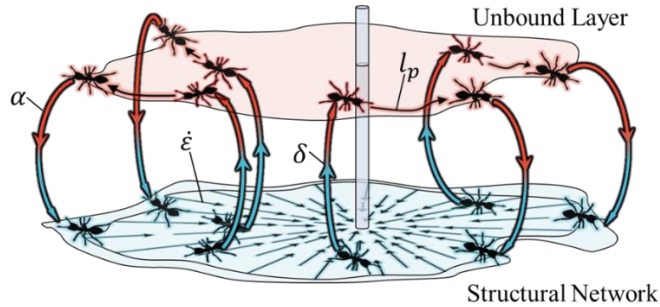

**Figure 4. Treadmilling of fire ant rafts:** Contraction of the raft at constant rate $\dot{\varepsilon}$ perpetually pulls ants in the structural raft network (blue) inwards. Structural ants unbind from the network at a flux rate of $\delta$ and become part of the active layer of ants (red). The freely active layer walks on top of the structural network with an approximate persistence length $l_p$ until they encounter the edge of the raft. Edge-encountering ants either bank off the edge of the raft or bind into the structural network at a flux rate of $\alpha$.

**Collective motion promotes protrusion growth:** Before examining the contributing factors to protrusion formation, we first characterized their characteristic growth rates and widths. Protrusions grow at an average measured rate of $\langle \dot{L} \rangle = 0.74 \pm 0.05 \; \ell \; \text{min}^{-1}$ with an average width of $\langle W \rangle = 5.85 \pm 0.06 \; \ell$ (**Fig. 5.A-B**) suggesting that the areal tip growth rate per unit edge length is on the order of $\langle \dot{L} \rangle \langle W \rangle = 4.33 \pm 0.08 \; \ell^2 \; \text{min}^{-1}$ or ~ 11 ants per minute (assuming a nominal structural raft density of $\rho_r \approx 0.3$ ants mm$^{-2}$). Normalizing this value by the approximate length of the leading tip edge (taken roughly as $\langle W \rangle$) we see that the average tip growth rate is roughly $\gamma_{tip} \approx 1.9$ binding events per minute per unit body length, which is an order of magnitude higher than that of the nominal raft ($\gamma \approx 0.29 \; \text{min}^{-1} \; \ell^{-1}$) we measured earlier. This disproportionate growth rate could either be due to a higher flux of surface ants to the tips of protrusion, a higher probability of surface ant deposition into the raft at these locations, or both. Whether the probability of edge parking varies by location is difficult to measure directly through experiments for two reasons. Firstly, defining the length scale over which an ant detects the edge is not easily quantified, and so identifying edge encounters is exceedingly difficult. Secondly, edge accumulation effects (26) ensure clustering of ants near the edges and image tracking individual ants in these regions becomes implausible due to indistinguishability of ants in close or overlapping proximity. As such, in the scope of this work we focus on identifying the existence and cause of heterogenous surface ant distribution and anisotropic flux on protrusions.

To discern the underlying contributions to excessive tip growth, we examined the trajectories of unbound ants on these features. We discovered that unbound ants on protrusions display a high degree of collective motion as characterized by their mean directional correlation (**Fig 5.C**). In fact, on protrusions, ants separated by more than $10 \; \ell$ exhibit statistically the same directional correlation ($c_{ij} \approx 0.25$) as ants separated by the contact length scale ($\sim 1\ell$) above which no correlation exists on the bulk of the rafts (**Fig 2.D**). The mean value of $c_{ij}$ across all separation distances when $\tau = 0$ s is $0.170 \pm 0.003$, which suggests that the ants are walking on average with a nominal separation angle of roughly 80°. This may seem like a large angular difference, but it generally suggests that ants are walking on average within the same quadrant of directional orientation and indicates a biased flux of ants moving in some direction on the protrusion. To confirm and identify this direction of bias we examine a multitude of measures for surface ants on a protrusion whose longitudinal axis is aligned with the $x$-axis of analysis. Firstly, we examine the metric tensor of ant velocity defined by $g^v = \langle \hat{v}_i \otimes \hat{v}_i \rangle_N$ where $\hat{v}_i$ represents the direction of motion of a single ant and the operator $\langle \rangle_N$ denotes taking the average over the sample size, $N$. The diagonal components of this tensor indicate the degree of motion projected onto the reference coordinate system where if $g^v_{11} > g^v_{22}$ then the $x$-compenents of velocity generally exceed the $y$-components. Indeed, we find that $g^v_{11} = 0.66$ while $g^v_{22} = 0.34$, suggesting that the ants are traveling faster along the longitudinal axis of the protrusion than the lateral axis. We find that $g^v_{12} = g^v_{21} = 0.01$ confirming that not only are the ants generally moving faster along the longitudinal direction than the lateral direction, but that the longitudinal direction is close to their principally fastest direction motion. The principal components of $g^v$ are displayed in **Fig. 5.D** to illustrate this alignment. To provide further visual transparency, both the 2D velocity distribution and isolated traffic of surface ants are presented in **Fig. 5.F** and **Fig 5.H**, respectively. Examining the 2D velocity distribution, the maximum longitudinal ($x$-axis) component exceeds that of the transvers ($y$-axis) component. Observing **Fig 5.H** (see **Movie S7** for better visual clarity), the ants on the protrusion generally move longitudinally, with little transverse motion. For reference, the metric tensor, velocity distribution and PIV-isolated surface traffic off a protrusion are depicted in **Fig. 5.E**, **Fig. 5.G** and **Fig. 5.I**, respectively and demonstrate that ants far from the edge of the raft move roughly isotropically ($g^v_{11} \approx g^v_{22} \approx 0.5$).

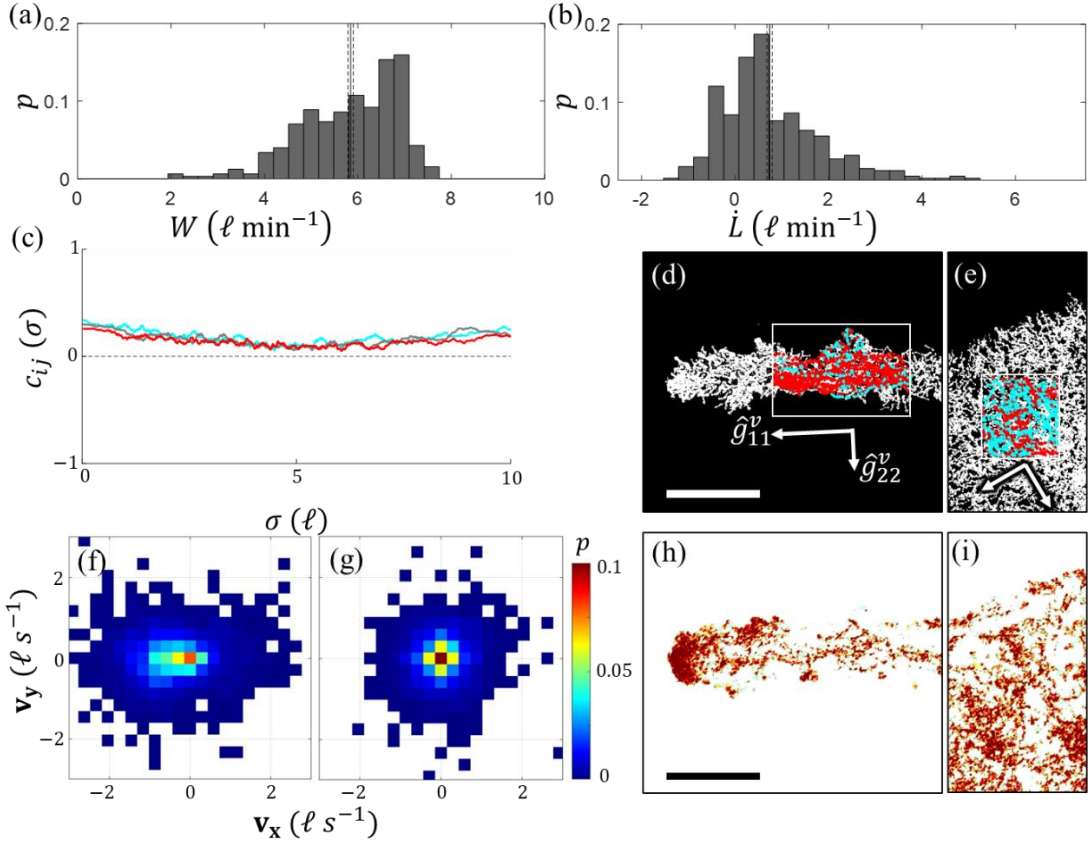

**Figure 5**. **Characterizing protrusion growth and collective motion:** **(A)** A distribution of 326 protrusion width ($W$) observations over 13 measured protrusions is displayed. The mean of width of all observations is represented by the vertical line with the dotted lines representing the range of standard error. **(B)** A distribution of 406 frame-to-frame protrusion growth rate ($\dot{L}$) observations over 13 protrusions is displayed. The mean of growth rate of all observations is represented by the vertical line with the dotted lines representing the range of standard error. **(C)** The moving average of $c_{ij}$ is plotted with respect to separation distance for $\tau = 0\,s$ (cyan), $\tau = 1\,s$ (grey), and $\tau = 10\,s$ (red). The moving average window was set to $1\,\ell$ to reduce noise for transparency. **(D-E)** All ant trajectories within a domain **(D)** on a protrusion and **(E)** on the bulk of a raft were manually image tracked. Ants moving left were overlaid with red dots and ants moving right were overlaid with cyan dots to illustrate the flux of ants during this time span. The principal directions of $\hat{g}^v$ are illustrated to display bias in movement. **(F-G)** The 2D velocity distributions of ants tracked in domains **(F)** on a protrusion and **(G)** on the bulk are displayed. **(H-I)** PIV was utilized to isolate the movement of surface ants on the raft **(H)** on a protrusion and **(I)** on the bulk to visually indicate direction of motion and clustering. All scale bars represent $10\,\ell$.

Although $g^v$ indicates that surface ants generally move longitudinally along protrusions, it reveals nothing about the sense of this movement (*i.e.*, whether movement primarily occurs from tip-to-base or base-to-tip). However, the positive value of $c_{ij}$ indicates that there is, in fact, a bias in directional flux (since ants moving in-line but opposite to one another will result in $c_{ij} = -1$). To identify the direction of this bias, we examine the 2D velocity distribution (**Fig. 5.F**) and observe that there is a slightly higher probability of ants moving up to $1\,\ell\,s^{-1}$ towards the tip than moving towards the base (*i.e.*, the distribution is skewed slightly left). To emphasize this flux illustratively, **Fig. 5.D** includes red points wherever an ant was recorded moving left ($\hat{v}_1 < 0$), and cyan points wherever an ant was recorded moving right ($\hat{v}_1 > 0$). Note that all ants within this domain and timeframe were recorded to eliminate selection bias. The velocity distribution and binary plot from

**Fig. 5.D** both indicate that within the recorded timeframe, most ants moved from the base-to-tip of the protrusion.

That the ants exhibit directional flow on the protrusions is consistent with studies on other systems of self-propelled particles and is attributable to the fact that ants on protrusions are effectively in channeled confinement in which the length scale of the confining dimensions (here the protrusion width $\langle W \rangle \sim 5\,\ell$) are less than that of ants' walking persistence length ($l_p \sim 20\,\ell$). Under such conditions, it is well-documented that self-propelled particles with volume exclusion and alignment interactions phase transition to unidirectional ordered motion (27–29). Given ants' short-range directional correlation and relatively long persistence length, it is thus fully expected that unidirectional laning will be commonplace on raft protrusions. Given the finite length of the protrusions, this flux results in clustering of ants at the dead-ended tip, which is visible in the isolated surface traffic from **Fig. 5.H**. It stands logically that even if the probability of edge deposition at the tips of protrusions equals that of deposition anywhere else along the edge of the raft, the higher frequency and duration of edge encounters (*i.e.,* clustering) of ants at the tip resulting from this anisotropic flux will contribute to a higher local growth rate. As such, it is reasonable to conclude that collective motion contributes to the runaway growth of instable protrusions.

**Discussion**

While collective motion of ants explains runaway protrusion growth, it does not explain what causes the breaks in local edge symmetry that initiate these features to begin with. The accumulation of self-propelled particles near confinement boundaries with higher local convex curvature is a well-documented effect (15, 26, 28). As expected, we also observe accumulation of surface ants in raft regions with initially higher edge curvature. Once initial convex breaks in edge symmetry were observed, it was often the case that protrusions eventually emanated from these regions (see **Movie S5** for example with isolated surface traffic). While this certainly provides one explanation for the initiation of proto-protrusions, there were also instances observed in which proto-protrusions visibly emerged in regions of low curvature (see **Movie S8** for example with isolated surface traffic). These appear due to clusters of ants seemingly spilling over the edge of the raft. This anecdotal evidence may indicate that there exists an anisotropic effective edge pressure whose magnitude is governed by factors such as the angle at which ants approach the raft edge or number of ants moving in unison. A comparable phenomena occurs in the generation of tether-like protrusions in lipid vesicles containing self-propelled Janus particles and the emergence of these protrusions is attributed to cooperative force generated by the particles when they move in synchronously (30). Remarkably, in these systems protrusions emerge spontaneously, requiring no specific stimuli such as external environmental gradients. Although we see no experimental evidence of long-range coupling, environmental gradients or centralized control driving the protrusions in ant rafts, it remains to be seen whether this phenomenon may be replicated in the absence of such stimuli. In future work, we aim to address these outstanding considerations and discern what are the local-scale properties that drive global shape evolution of ant rafts or comparable systems through *in silico* experimentation using an agent-based numerical model.

Our findings confirm those of earlier works which posit that ant raft expansion is caused by the deposition of free, active surface ants into a structural raft network that floats on the water. Yet despite this outwards expansion, we find that the structural networks of ant rafts undergo contraction at a constant strain rate thus demonstrating that transport of ants occurs in both phases of the raft, albeit at different time scales. These expansion and contraction dynamics loosely balance one another and culminate in a perpetual treadmilling of ant rafts that is prerequisite to the significant shape evolution we observe over hour-long timescales. Fascinatingly, this treadmilling mirrors that of cellular cytoskeletons (2) at the nanometer length scale. Just as fire ants utilize treadmilling to generate tether-like protrusions which their colonies can then use to escape flooded environments, cytoskeletal systems leverage treadmilling to evolve their global membrane shape and facilitate motility (3). Despite their vastly disparate length-scales, these two biological systems exhibit remarkably convergent phenotypical behavior in which the emergent properties of the overall systems are dictated by the binding and unbinding dynamics of constituents at the local level. As such, they may provide a natural model for researchers to mimic in the production of synthetic materials or swarm robotics that achieve comparable functions.

Ultimately, our study characterized two novel behaviors of fire ant rafts that transcend length scales treadmilling dynamics that closely mirror those of cytoskeletons (3) and protrusion dynamics whose origins and evolution elicit comparisons to fingering of synthetic cell-like vesicles containing Janus particles (28). As a result of treadmilling, ant rafts are always out-of-equilibrium, allowing them to continually change their shape and explore their environments. This non-equilibrium state yields the emergence of unstable protrusions, commonly in locations of initially higher edge curvature, that quickly extend from the raft due to the unidirectional collective motion of surface ants confined on these formations. While the spontaneity and local behaviors that impact global raft shape remain to be examined in future work, ant rafts prove to be an exemplary macroscopic system through which to examine the dynamics of both condensed and free active matter and elucidate the physical mechanisms that drive emergent functional behaviors.

**Materials and Methods**

**Experimental Design:** To conduct our experiments, we collected on the order of 3,000-10,000 worker ants and placed them into a container of water in which an acrylic rod protruded from vertically from the water-line. The ants enveloped and nucleated to the acrylic rod, thereby anchoring them to a still reference frame. Rods with both 6.3 mm and 15.8 mm diameters were used in experiments. Additionally, both Fluon© and talcum powder were applied to some rods to prevent ants from climbing on them while others were left bare. Finally, the degree to which rods protruded from the water's surface was varied from roughly 1 cm to 15 cm. Under all boundary conditions, ant rafts displayed treadmilling in which they expanded outwards due to surface ant deposition and retracted inwards due to structural network reconfiguration. Whether treadmilling occurs only in the presence of an anchoring rod also remains to be seen, but the treadmilling dynamics presented herein occur far from this boundary condition. It is also still unclear whether the rods' presence effected the emergence of protrusions as they grew given a wide range of boundary conditions (*e.g.*, rod diameters, rod heights, and non-stick coatings). Observationally, the primary factor that appeared to govern their appearance was the activity level of ants.

In all experiments allowed to run sufficiently long, the surface ants eventually assumed an inactive state in which edge deposition ceased and the surface ants clustered around the rod. As such, protrusion phenomena were all captured within the first several hours of experimentation. However, in the scope of this work, an intermediate phase is sometimes presented in which some inactive surface ants clustered to the rods while other active surface ants contributed to outwards expansion. Since this former group of non-participating ants represented a tertiary phase that did not contribute to treadmilling dynamics, they were neglected in the characterization of surface ants, and raft contraction was only measured in regions where inactive, rod-clustering ants did not obstruct the view of the structural network. Regardless, the net effect of this clustering was a reduction in overall raft area over long timescales, due to a decrease in edge binding-driven expansion; therby preventing indefinite observation of the concerning phenomena. To ensure ample ant activity upon initiation of experiments, the air temperature in the room was monitored and controlled to remain between 20° and 24° C. The water temperature was also monitored, albeit not controlled and remained between 17.0° and 19.9° C throughout our experiments.

Charge-coupled devices were placed above the ant rafts to capture real-time and time-lapsed footage from a top view. Real-time footage was captured roughly every 10 minutes throughout the duration of the experiments to ensure a representative temporal sampling. Time-lapse footage was captured throughout the entirety of some experiments to measure structural contraction.

**Estimating ant packing fraction:** Video footage was imported into and processed using ImageJ (31–33). All data post-processing was achieved through MATLAB 2019b (34). Length scale calibration was always achieved by measuring a reference ruler placed in-frame, as close as possible and parallel to the water line. Planar raft density, $\rho_r = 0.304 \pm 0.018$ ants mm$^{-2}$, was measured by drawing rectangular regions of interest of known area and counting the number of raft ants residing within them. To normalize the binding and unbinding dynamics of ants into and out of the structural raft later, we define the areal occupancy of one structural ant in the raft network to be $\zeta^2 = \rho_r^{-1} \approx 3.2$ mm$^2$ per ant.

The surface layer density, $\rho_s = 0.072 \pm 0.006$ ants mm$^{-2}$, was found by counting the surface ants in the same regions of interest as the structural ants. Surface ants were distinguished from raft ants by toggling back and forth between adjacent frames in our video to detect which ants were moving (surface ants). The relative packing fraction of surface ants is $\phi \approx 0.24$ surface ants per structural ant, suggesting that in the regions measured, there is roughly 1 surface ant for every 4 structural ants. This value may in fact be higher due to the difficulty of reliably measuring $\rho_s$ where low-contrast, high-density clusters occurred, however it is apparent that the majority of ants in the rafts far from the anchoring rods at any given time occupied the structural network.

**Free ant speed:** Mean surface ant speed, $v_0$, was measured using manually image tracked data in ImageJ. Real-time, top view footage was spatially partitioned into distinct regions of interest in which any ant visibly moving on the surface had its position tracked, frame-to-frame (see sample of ants with overlaid trackers in **Movie S9**). For consistency, the petiole of each ant was always the anatomical position tracked. We tracked every ant that entered or exited the region of interest within the duration of the video to accurately represent the population and prevent selection bias. Note that the data is inclusive of ants that were visibly on the surface (*i.e.*, unbound and active), yet were not traveling, due to stopping upon ant-to-ant contact or self-cleaning. A snapshot of image-tracked ants with their streamlines plotted and the full distribution of 19,970 discrete frame-to-frame observations are shown in **Fig. S1.A** and **Fig. S1.B**, respectively. The full 2D trajectory space of all frame-to-frame observations across two experiments is presented in **Fig S2.A-B**. From **Fig S1.B**, we see that the mean observed surface ant speed, $v_0$ is $0.59 \pm 0.01$ $\ell$ s$^{-1}$. Note that at any given moment, there is a large probability ($\sim 17\%$) of a surface ant moving at less than $0.08$ $\ell$ s$^{-1}$ (or $0.14$ mm s$^{-1}$). As this corresponds to roughly the error length of 1 pixel in manual image tracking, these ants are effectively stationary due to things such as self-cleaning or entanglement with clusters of neighboring surface ants. This highlights the importance of including non-moving surface ants in the measurements, as they constitute a large percentage of the surface ant population at any given moment, and pausing is often a precursor to cluster formation (15). These temporarily paused ants also explain the bouts sub-diffusive behavior observed later when the mean-square displacement is examined.

**Free ant persistence length:** With the trajectory data available from manual tracking, we were able to estimate the persistence length, $l_p$, of surface ant trajectories. Persistence length is defined as the travel distance or contour length, $l_c$, over which correlation between an ants' trajectory at reference time $t_0$ and some future time $t$ is lost (35). To find $l_p$, first, we calculated the ensemble average of the trajectory correlation function, $C(v_i, t)$, which may be written as:

$$C(\hat{a}_i, t) = \langle \hat{a}_{i,1} \cdot \hat{a}_{i,k} \rangle_{N_{it}}, \qquad (1)$$

where $k$ represents the $k^{th}$ iteration up to time $t$, $\hat{d}_{i,1}$ is the $i^{th}$ particle's displacement direction at the first iteration of interest, $\hat{d}_{i,k}$ is the same particle's displacement direction at the $k^{th}$ iteration, and $\langle\rangle_{N_{it}}$ denotes taking the ensemble average over all $N_{it}$ iterations. Note that at time $t = 0$, $C = 1$ by definition. Averaged over many randomly walking particles, $C$ will decay exponentially in time as a particle's trajectory becomes less correlated with its original value (35). We define the contour length, $l_c$ traveled by a given particle with respect to time as the sum of the distances traveled within each step:

$$l_c = \sum_{k=1}^{N_{it}} d_{i,k}. \qquad (2)$$

Plotting $C$ against $l_c$ gives us the correlation function with respect to distance traveled. From this plot, we use MATLAB R2019b (34) to fit a curve to the data of the form:

$$C = \exp\left[-\frac{l_c}{l_p}\right]. \qquad (3)$$

The extended data set of $C$ is shown in **Fig S2.B-C**. Via this method it is clear that $C$ does not decay exponentially and the ants do not walk completely randomly. However, this method is useful in identifying the order of $l_p$, which was found to be approximately $20.5 \pm 5.7\ \ell$ and is comparable to the confining raft dimensions. This ensures that active ants regularly encounter the edges of the raft as self-propelled particles with persistence lengths comparable to their confines' dimensions have been shown to do (14,15). Regarding data selection, to avoid inclusion of edge effects in measurements we excluded image-tracked ants that encountered the edge. Additionally, if the amount of data for a given trajectory was insufficient such that $R^2 \leq 0.5$ for our exponential fit this data was excluded from our sampling and considered unreliable for extrapolating $l_p$.

**Free ant mean square displacement:** The mean square displacement, $\langle x^2 \rangle$, of surface ants was calculated according to:

$$\langle x^2 \rangle = \langle |x_i(t+\tau) - x_i(t)|^2 \rangle_{N_p} \qquad (4)$$

where $\tau$ is the time interval over which displacement is measured and $\langle\rangle_{N_p}$ denotes taking the ensemble average over all $N_p$ particles. The extended data of $x^2$ prior to ensemble averaging is depicted in **Fig S2.E-F** for two separate experiments to illustrate the variance in diffusion. Also note that there are instances in which individual ants exhibit momentarily sub-diffusive behavior, likely caused by their propensity to stop to clean themselves or due to collision with other ants and clusters.

**Free ant normalized order parameter:** The normalized velocity order parameter, $\varphi$, of surface ants was calculated according to:

$$\varphi = \frac{\langle v(t) \rangle_N}{\langle |v(t)| \rangle_N}, \qquad (5)$$

where $v$ is an ant's velocity and the operator $\langle\rangle_N$ denotes taking the average over all $N$ ants occupying the designated region of interest. This parameter is one when self-propelled particles are perfectly aligned in their trajectories, and approaches zero when the trajectories are completely random (36). We subdivided the regions of interest into increasingly smaller domains and tracked $\varphi$ against this domain size, $L$. Domains containing only 1 ant within a given set of frames were excluded, as these would exhibit $\varphi = 1$, by default.

**Visually isolating surface traffic:** PIV was conducted via PIVlab (24, 25) on a variety of raft geometries to isolate the motion of surface ants despite the low contrast background (of the structural raft ants). To ensure overlap between a given ant with itself in adjacent frames, the timestep between frames used for this purpose was 0.67 s or less. The resolution of PIV was set to an arbitrarily small fraction of 1 $\ell$. Using this resolution, the vibrational noise due to movement of surface ants as they not only locomoted, but also turned, moved their legs, moved their antennae, *etc.* was captured while excluding the relatively inactive structural ants whose global speed of contraction was orders of magnitude lower than the speed of the surface ants. To additionally remove the effects of raft motion due to factors such as raft rotation, we subtracted the local mean velocity taken arbitrarily as the spatial moving average over a window of 2 $\ell$. This window was adjusted until qualitative clarity in the trajectory of surface ants was achieved. Finally, 4 s of footage are overlaid onto each frame such that an ant moving at the average measured speed ($v \sim$ 0.6 $\ell\,\text{s}^{-1}$) will be displayed over a distance greater than 2 $\ell$, thereby visually illuminating the direction of motion to some degree and emphasizing clustering. Snapshots of various ant rafts geometries before and after durations on the order of $10^2$ s are presented in **Fig S3** to illustrate the distribution of surface ants driving raft shape evolution.

Although qualitative in nature, these figures illustrate the ubiquitous clustering of surface ants, especially near convex regions of edge curvature, and how this clustering generally leads to protrusion initiation. **Fig S3.A-B** illustrates the initiation of a protrusion in a convex region of raft edge curvature, which was the most common mode of growth initiation. **Fig S3.C-D** and **SI Fig 3.E-F** illustrate how clustering is exacerbated at the tips of both relatively young and mature protrusions alike, respectively, which leads to runaway growth. Contrasting **Fig S3.A-B**, **Fig S3.G-H** illustrates the spontaneous initiation of a protrusion at a region of relatively low edge curvature, seemingly due to the synchronized motion of cluster of ants that stochastically approaches the edge approximately normal to its length. This indicates that not only raft edge curvature, but also cooperative motion of surface ants or clusters leads to higher rates of edge deposition and suggests some degree of anisotropic effective edge pressure. For additional visual clarity and to perceive the motion of these ants, as well as the original video footage from which these figures were taken, also see **Movies S5-S8**.

**Manually measuring structural contraction:** We took $\dot{\varepsilon}$ as the radial strain rate of raft ants towards the stationary acrylic rod using manual image tracking via ImageJ. This was done since the acrylic rod was a stationary reference frame with respect to our charge-coupled devices. In order to manually track the contractile strain rate, $\dot{\varepsilon}$, we selected a set of ants along the perimeter of the raft at some reference time and used ImageJ's manual image tracking plugin to track this set of ants frame-by-frame as their position shifted inwards towards the center of the raft with time. We did this for several different starting frames across three experiments. A raft with the contours traced by distinct sets of these points (shown in different colors) are presented in **Movie S3**. Given the displacement of each ant across each set of frames, we can calculate their velocities, $\boldsymbol{v}$. To exclude the effects of raft spin about the acrylic rod, we projected the velocity of each ant tracked onto their unit vectors directed towards the center rod, $\widehat{\boldsymbol{R}}$, according to:

$$\dot{R} = \boldsymbol{v} \cdot \widehat{\boldsymbol{R}}, \qquad (6)$$

thus isolating components of velocity towards the still reference frame, $\dot{R}$. To reduce temporal noise, we took temporal moving averages of $\dot{R}$ over 3-minute intervals (less than 3 % over our shortest experiment duration). Plotting $\dot{R}$ with respect to distance from the rod $R$, we estimated $\dot{\varepsilon}$ as the slope of the least-squares regression line fits to the data:

$$\dot{\varepsilon} \approx -\dot{R}R^{-1}, \qquad (7)$$

where $R$ is the distance from the center of the rod to the point where velocity was measured. Via manual image tracking, we found that $\dot{\varepsilon}$ is anywhere from 0.6 to 4.4 % min$^{-1}$, depending on the experiment and level of activity. The complete ensemble of all observations is plotted for three separate experiments in **SI Fig 4.A-C**. with respect to $\dot{\varepsilon}$. The lack of significant correlation between $\dot{\varepsilon}$ and $R$ ($R^2 = 0.08, 0.17$, and $0.00$, for linear least-squares regression fits, respectively) initially suggests that the strain rate is, at least, radially constant within ant rafts.

However, given the high degree of noise, taking the moving average and plotting the ensemble data from all three experiments on one plot (**Fig S4.D**), we see that $\dot{\varepsilon}$ may decrease at distances less than $10\ \ell$. To ensure that this reduction in $\dot{\varepsilon}$ is not merely attributable to differences in experiments (Experiments 2 and 3 are both smaller than Experiment 1), we also plot the moving average of the entire ensemble of data from each of the three experiments separately (**Fig S4.E**) and there remains a marked decrease in $\dot{\varepsilon}$ when $R < 10\ \ell$. While the exact cause of this trend is unknown, it may be due to a higher surface packing fraction of surface ants near the center of each raft due to clustering of inactive ants in this region. This may inhibit the rate of structural ant unbinding locally – a significant mechanism identified in this work as driving or permitting raft contraction. Logically, the structural ants with surface ants residing on top of them will be less likely to unbind from the network and move into the surface layer. Thus, in regions of higher surface density, the raft may be unable to contract without a significant increase in structural network density. In experiments 1 and 2, accumulation of surface ants is visibly apparent near the base of the acrylic rod (see **Fig S5.A-B**) over long timescales. In experiment 3 (see **Fig. S5.C**), while there is no visible accumulation of surface ants at the base of the rod in the time lapses, there is a notable repository of ants at the tip of the acrylic rod. Any flux into and out of this repository may result in a high concentration of surface traffic at the base of the acrylic rod, thus diminishing the local unbinding rate and contraction rate. We also examine the evolution of the average $\dot{\varepsilon}$ for each experiment (**Fig 4.F**) and see no discernable trend in average $\dot{\varepsilon}$ with respect to time. This suggests that the strain rate, while varying greatly in time, was neither consistently increasing nor decreasing throughout our experiments.

**Measuring structural contraction with PIV:** PIV was conducted using PIVlab (24, 25) on a continuous region of interest within the largest of experimental rafts over a 13 minute duration. A vector field within the designated region of interest was produced and velocities exceeding a certain magnitude ($|v| > 0.1\ \ell\ \text{s}^{-1}$) were then cropped from the data, as these data points represented noise due to surface ant traffic. The cropped data was then used to calculate $\dot{R}$ according to (**6**) and $\dot{\varepsilon}$ according to (**7**). For visual clarity, the vector field presented visually in **Fig. 3.D** is the temporal averaged vector field over the entire 13 minute analysis duration.

**Measuring the structural unbinding rate:** Structural ants unbinding from the raft network into the surface traffic layer was confirmed as the cause of inwards contraction through visual observation (see **Movie S4** for example) and explains the uniformity of the contraction rate throughout the bulk of the raft. Strain rate, $\dot{\varepsilon}$, may be leveraged to analytically determine unparking rate, $\delta$. We found that the raft contracts at a roughly constant strain rate, $\dot{\varepsilon}$ and that this rate roughly predicted the rate of areal contraction according to $A_r = A_r^0 e^{-2\dot{\varepsilon}t}$, where $A_r$ is the area circumscribed by a single set of structural ants that were originally at the raft perimeter and are tracked in time. To find the rate change in circumscribed area, $\dot{A}_r$, we take the time derivative as:

$$\dot{A}_r = -2A_r^0 \dot{\varepsilon} \exp(-2\dot{\varepsilon}t). \quad (8)$$

Assuming raft structural density is conserved, we may then take the normalized unparking rate as $\delta = \dot{n}_u/\langle A_r \rangle$, where $\langle A_r \rangle$ is the average retraction area between adjacent frames and $\dot{n}_u$ is the absolute unparking rate given by:

$$\dot{n}_u(t) = \dot{A}_r(t)\rho_r. \quad (9)$$

Substituting (**8**) into (**9**), and then normalizing by $A_r$ gives $\delta$ as:

$$\delta = -2\rho_r \dot{\varepsilon} \left(\frac{A_r^0}{A_r}\right) e^{-2\dot{\varepsilon}t}. \quad (10)$$

Therefore, at any given reference time (in which $t = 0$ and $A_r = A_r^0$) we are left with a nominal $\delta$ of $-2\rho_r\dot{\varepsilon}$. Given the wide range of $\dot{\varepsilon} \in [0.006, 0.045]$ min$^{-1}$ across experiments, we calculate the unparking rate as anywhere from 1% to 9% of raft ants (or $\zeta^{-2}$) per minute, which is consistent with the values in **Fig S5.D-F**. The unparking rate, $\delta$, may also be calculated experimentally, through areal image analysis. Given the manual image tracking data, we may take the areal decay rate of the raft to be the rate change of the area circumscribed by the set of initial edge ants, $A_r$. This rate change, $\dot{A}_r$, is given by:

$$\dot{A}_r = \frac{[A_r(t) - A_r(t-\Delta t)]}{\Delta t}, \quad (11)$$

where $\Delta t$ is the duration between adjacent frames. If we assume $\rho_r$ is roughly constant, then the rate of unparking events may be calculated simply as $\dot{n}_u = \dot{A}_r/\rho_r$. Recall that the normalized unparking rate is defined as $\delta = \dot{n}_u/\langle A_r \rangle$. The circumscribed area, $A_r$, is outlined for snapshots of three different experiments in red in **Fig S5.A-C**. The values of $\delta$ calculated according to this latter method and corresponding to these three experiments are plotted with respect to time in **Fig S5.D-F**. While there exists a strong degree of noise, values of this normalized unparking rate are consistently on the order of $\sim 10^{-2}$ unparking events per raft ant per minute, which are consistent with those values we may analytically calculate from the range of $\dot{\varepsilon}$.

**Measuring the free ant binding rate:** Surface ants binding into the edge of the raft was confirmed as the cause of outwards expansion through visual observation (see **Movie S2** for example). The areal growth rate of the raft was measured via manual image tracking of image stacks in ImageJ. Since the raft is constantly retracting, we calculate the updated growth zone area, $A_g$, at time $t$ as the difference in total raft area, $A$, and the traced raft area, $A_r$. Snapshots from three separate experiments **Fig S5.A-C** show the newly deposited structural ants (occupying area $A_g$) shaded in cyan. Thus, the rate change in $A_g$ is taken as:

$$\dot{A}_g = \frac{A_g(t) - A_g(t-\Delta t)}{\Delta t}, \quad (12)$$

Again, assuming $\rho_r$ is roughly constant, we may calculate rate of parking events as:

$$\dot{n}_p = \frac{\dot{A}_g}{\rho_r}, \quad (13)$$

Normalizing $\dot{n}_p$ by the average circumscribed raft area within this time duration, $\langle A \rangle$, within this time span gives the structural binding rate $\alpha$ as:

$$\alpha = \frac{\dot{n}_p}{\langle A \rangle}. \quad (14)$$

$\alpha$ may also be derived from the following balance equation:

$$\dot{N}_s = (\delta - \alpha)A, \quad (15)$$

where $\dot{N}_s$ is the rate change in the number of surface ants. Recognizing that $N_s = A\rho_s$, where $\rho_s$ is the surface ant density (assumed constant), we may write:

$$\dot{N}_s = \dot{A}\rho_s. \quad (16)$$

Substituting **(16)** into **(15)** and solving for $\alpha$, we get:

$$\alpha = \delta - \frac{\dot{A}}{A}\rho_s, \quad (17)$$

which was used to calculate the $\alpha$ in **Fig S5.D-F**. Again, while the frame-to-frame rate exhibits high noise, via both methods, we found that $\alpha$ is on the same order as $\delta$ ($\sim 0.02$ parks or unparks per minute per raft ant). Note that for all three experiments, neither $\alpha$ nor $\delta$ change according to any trend in time. This confirms that for all experiments, the raft area, $A$, was roughly conserved, or its rate of change was roughly conserved.

**Measuring protrusion growth rates:** Observed protrusion growth rates were measured using ImageJ. Since the raft is perpetually contracting, it is not sufficient to measure the change in overall protrusion length to determine the growth rate due to expansion. Instead, a set of two reference ants occupying the structural network layer on opposite sides of the protrusions were manually image tracked across each frame. Then the relative contour length, $L$, between the mean position of these two reference ants and the tip of the protrusion was measured as reference length $L$. A single growth rate observation was then taken as $\dot{L} = [L(t + \tau) - L(t)]/\tau$ where $\tau$ is the time between adjacent frames. 406 observations from 13 distinct protrusions were measured and used to calculate the mean growth rate, $\langle \dot{L} \rangle = 0.74 \pm 0.05 \, \ell \, \text{min}^{-1}$. To evenly weight measurements from each protrusion, $\tau$ was held constant across samples and every protrusion's growth rate was measured for the full duration of its recorded lifetime. To ensure that this method of calculating mean growth rate provided the right order of magnitude, we additionally calculated the slope ($dL/dt$) of the least-square linear regression fits to the data from each protrusion and found that growth was consistently on the order of $\sim 1 \, \ell \, \text{min}^{-1}$ indicating that the protrusion grew at around one body length per minute. Snapshots from a single sample of growth measurement are presented in **Fig S6** along with the data of $L$ plotted with respect to time.

**Estimating protrusion widths:** Mean protrusion width $W$ was approximate by circumscribing the area of protrusions $A_p$ and then tracing their contour along the length, $L_c$. $W$ was then taken as $W \approx A_p/L_c$

**Data and materials availability:** The data supporting the findings of this publication are available from the corresponding authors upon reasonable request.

# Acknowledgments

**Funding:** Franck J. Vernerey gratefully acknowledges the support of the National Science Foundation under Award No. 1761918. The content is solely the responsibility of the authors and does not necessarily represent the official views of the National Science Foundation.

**Supplementary materials**

**Figures:**

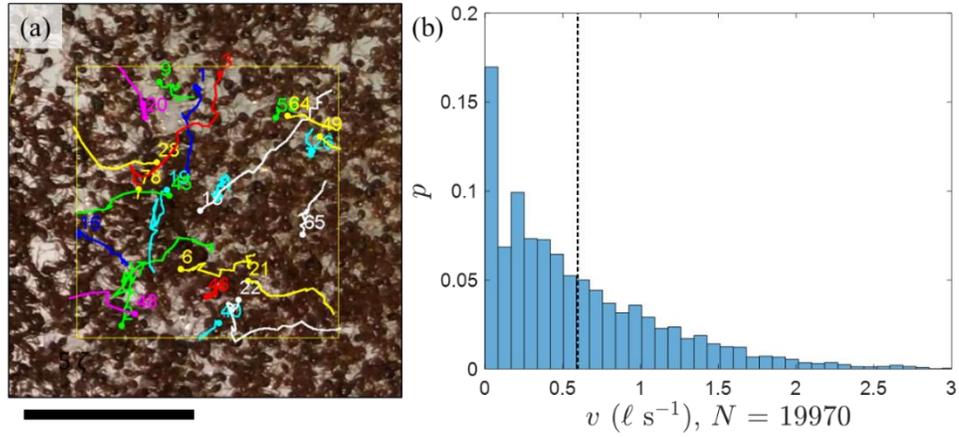

**Fig. S1. Extended ant speed data:** (**A**) A sample snapshot of an ant raft overlaid with manually image tracked trajectories (scale bar represents 5 $\ell$). (**B**) The probability distribution of surface ant speed is shown for a sample size, $N = 19{,}970$ discrete observations. The mean ant speed over all observations, $v_0 = 0.596 \pm 0.004 \; \ell \, s^{-1}$, is plotted with a dotted line.

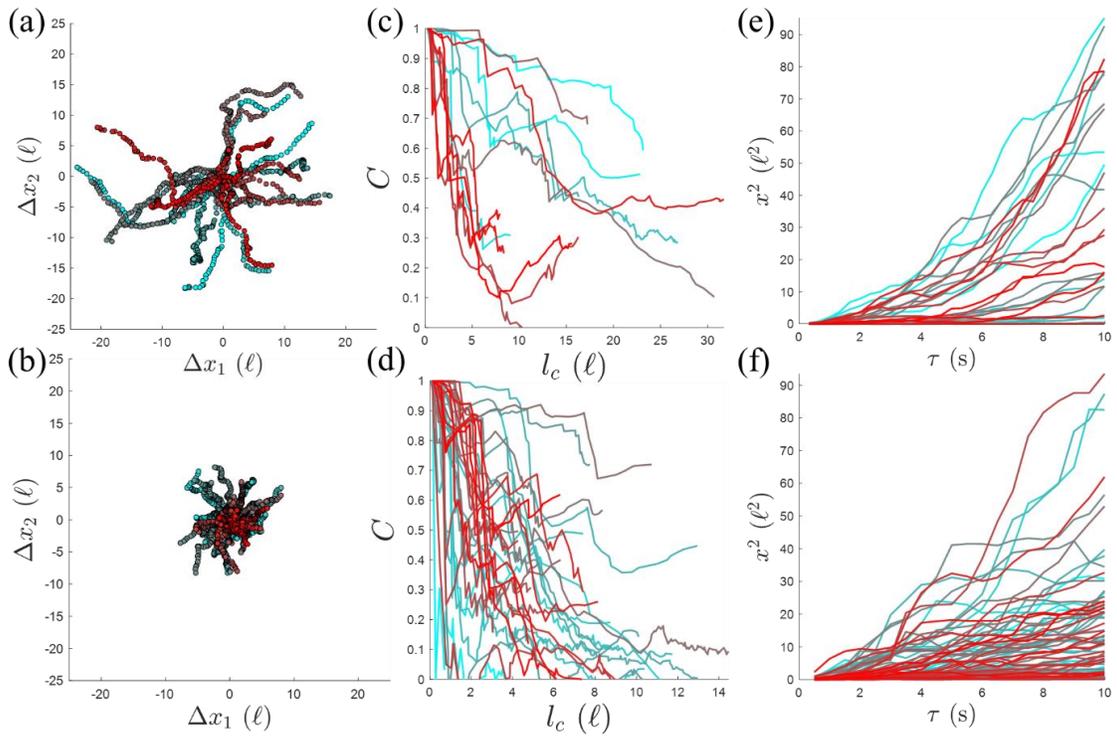

**Fig. S2. Extended trajectory analysis data: (A-B)** The trajectory space of all ants measured in two separate domains on the bulk (*i.e.,* without edge-encounters) of two separate ant rafts is displayed with each color representing a different ant. Note that differences in the domain size account for the difference in the maximum trajectory contour lengths measured between the two experiments. **(C-D)** $C(l_c)$ is plotted for all ants tracked displayed in **(A)** and **(B)**. The decay in correlation for any given ant may be approximated according to $C = e^{-l_c/l_p}$, however this method may only reliably approximate the order of ant persistence length and not a precise value. **(E-F)** $x^2$ is plotted with respect to $\tau$ for the full set of ants tracked in **(A)** and **(B)**. Anomalous diffusion is apparent from the non-linearity of these data sets and explains the difficulty in measuring a given ants persistence length. Ants sometimes alternate between phases of super and sub-diffusive behavior as illustrated by the divergence and plateauing of the slopes within different regimes. The data in **(C-F)** is plotted continuously purely for illustrative purposes. Full discrete data sets may be made available upon reasonable request.

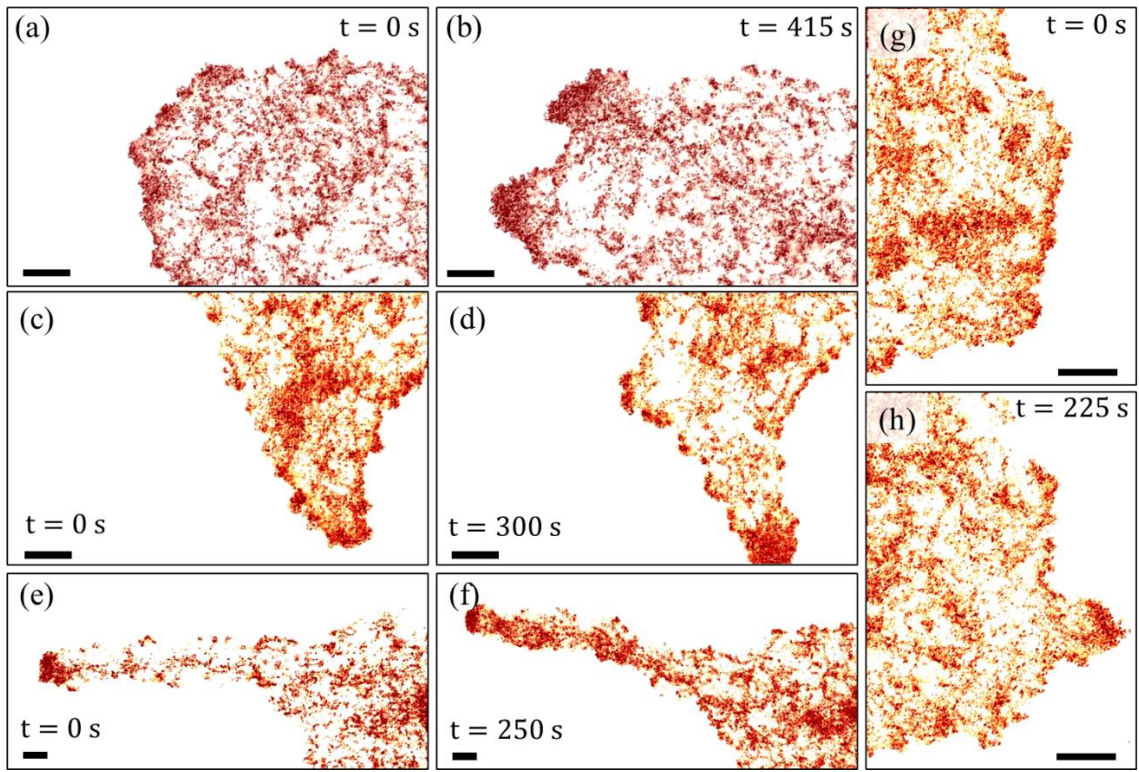

**Fig. S3. PIV-isolated surface traffic:** These images depict the isolated surface traffic on ant rafts in orange or red. (**A-B**) The initiation of two adjacent protrusions due to edge clustering is displayed in a region of locally higher convex edge curvature, thus demonstrating one common way in which protrusions were observed forming. (**C-D**) The growth of a relatively young, and highly tapered protrusion is depicted. In such a case, the raft geometry effectively funnels surface ants to the tip where they cluster and promote runaway growth and the formation of tether-like structures like that depicted in (**E-F**). (**E-F**) Tip clustering and runaway growth were often sustained in longer, more mature protrusions without additional tapering. (**G-H**) The initiation of a protrusion is depicted in a region of raft edge with a relatively low edge curvature, indicating that while convex edge curvature may facilitate asymmetric growth, there are other factors such as cooperative motion of surface ants that can lead to breaks in symmetry. All scale bars represent 5 $\ell$.

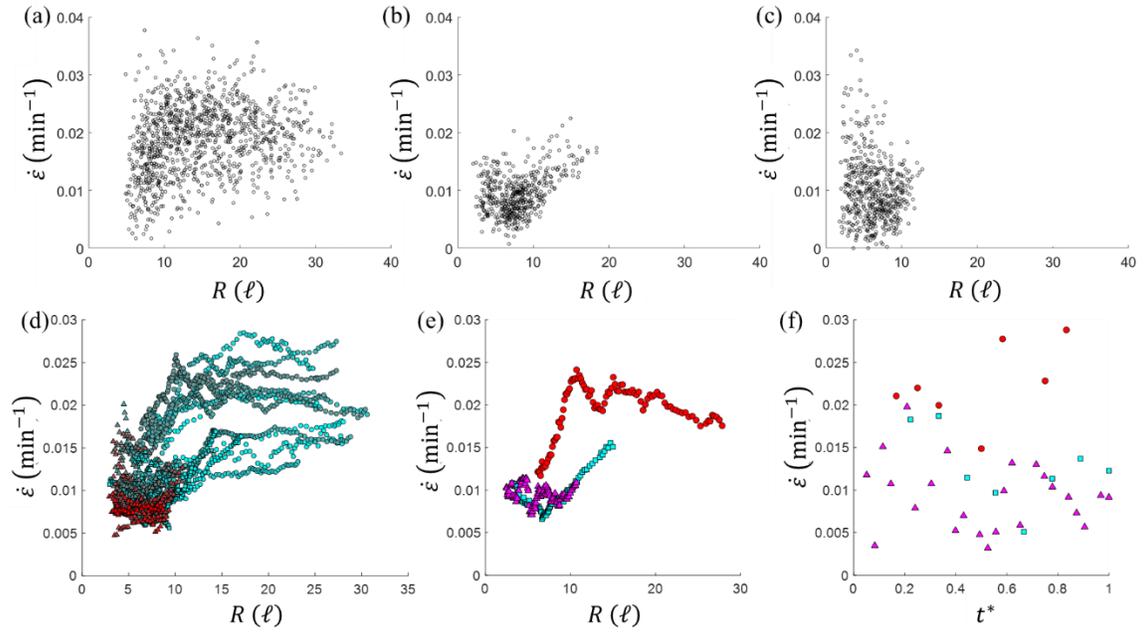

**Fig. S4. Extended structural contraction data: (A-C)** The full set of observations of $\dot{\varepsilon}$ is plotted with respect to $R$ for the largest to smallest (**A-C**, respectively) raft experiments. (**D**) The entire dataset of $\dot{\varepsilon}$ is plotted with respect to $R$. Circles, squares and triangles represent data sets from the largest to smallest experiments conducted, respectively. Different colors denote different samples across all experiments. (**E**) The moving average of $\dot{\varepsilon}$ is plotted with respect to $R$ for each of three experiments. (**F**) $\dot{\varepsilon}$ is plotted with respect to normalized time, $t^* = t/t_{max}$ for each of three experiments. (**D-E**) Red circles, cyan squares and magenta triangles represent the largest, intermediate, and smallest experiments, respectively. In (**E**) and (**F**), red circles represent experiment 1 from (**A**), cyan squares represent experiment 2 from (**B**) and magenta triangles represent experiment 3 from (**C**).

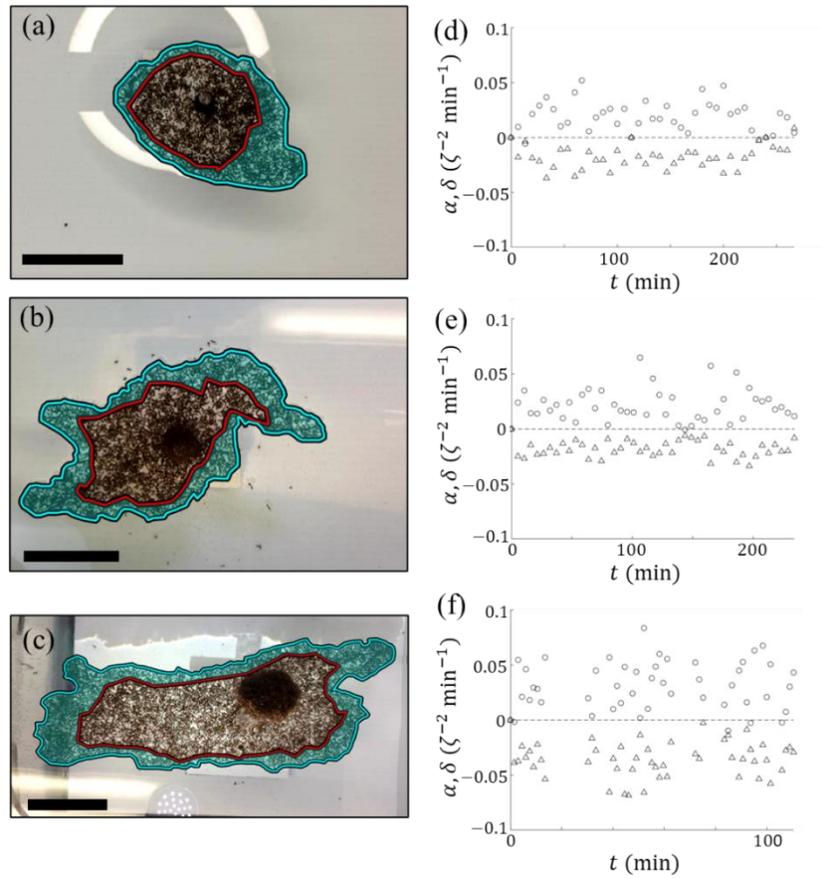

**Fig. S5. Extended treadmilling dynamics data: (a-c)** Snapshots of 3 experiments of varying sizes are shown. The red contour represents structural ants who were originally at the raft perimeter at reference time, $t_0$, and flowed inwards in time. This contour circumscribes the area $A_r$. The region shaded in cyan represents the growth zone, $A_g$ of newly deposited structural ants into the edge of the raft network since time $t_0$. **(d-f)** The structural binding ($\alpha$ as circles) and unbinding ($\delta$ as triangles) rates are plotted with respect to time in units of events per $\zeta^2$ per min where $\zeta = \rho_r^{-0.5}$ is the length scale that one square structural ant occupies. In these units, $\alpha$ and $\delta$ may be thought of the as the instantaneous areal expansion and reduction rates of the raft, respectively, and if the two values are approximately equal then raft area will be loosely conserved. Gaps in the data of **(f)** correspond to time spans during which portions of the raft exited the visible frame and from which data could not be accurately taken.

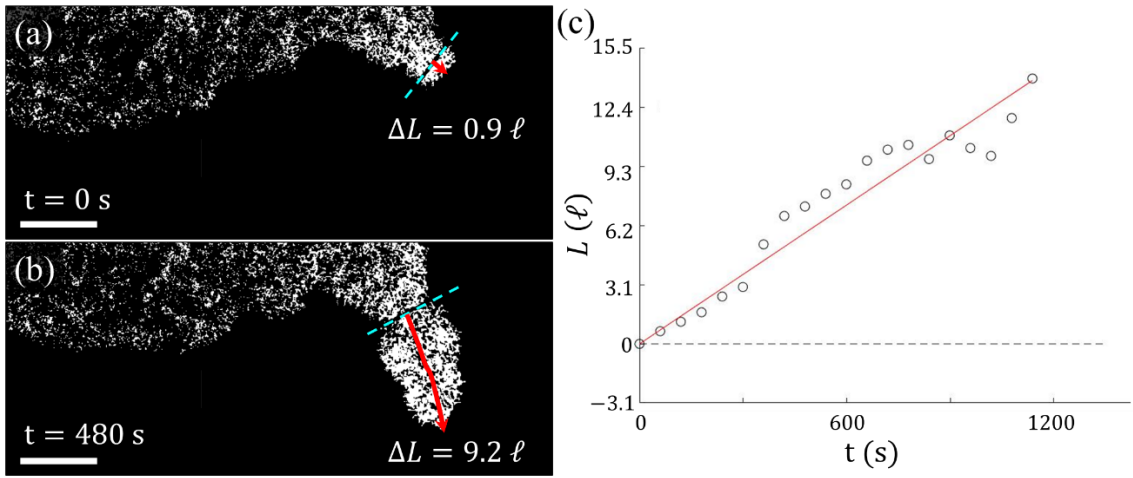

**Fig. S6. Protrusion growth measurement sample: (a-b)** A binary image of a raft (white represents the raft) at the **(a)** start and **(b)** end of a 480 s duration depicts the growth of a protrusion from the bottom right edge. The protrusion's length $L$ (red arrow) was measured from a set of reference ants image tracked in time (dotted cyan line) to the approximate tip of the growth. **(c)** The sample growth rate was taken as the slope of the linear regression fit to the plot of $L$ against time, $t$. Here, $\dot{L} \approx 1\, \ell\, \text{min}^{-1}$ ($R^2 = 0.93$).

**Movies:**

**Movie S1 – Ants as Biphasic Matter** is real-time footage of the edge of an ant raft that visually illustrates the contrast between the structural raft network, which is relatively static, and the freely active layer of surface ants in which the ants are dispersed and highly motile. The scale bar represents 5 $\ell$.

**Movie S2 – Parking Examples** illustrates close-up footage of the edge of an ant raft in which three distinct, unbound ants walk up to the edge of the raft. These ants are then pressured (either pushed or stepped on by neighboring unbound ants) until they park and become part of the raft-bound layer. These ants are denoted with overlaid dots labelled "1", "2" and "3".

**Movie S3 – Raft Treadmilling** illustrates traced sets of bounded raft ants as they flow inwards in time. These bound ants that were originally on the perimeters of the rafts, flow inwards. Different sets of ants are traced by distinct colors. A new set of ants was tracked roughly every 22 minutes to visually illustrate raft contraction. This video also clearly exhibits the formation of newly deposited raft area outside of each image tracked contour, indicative of edge parking by surface ants. Together, these competing mechanisms give rise to radial treadmilling.

**Movie S4 – Unparking Examples** illustrates close-up footage of an ant raft in which two distinct ants originally bound in the raft unpark and move into the unbound surface layer. These ants are denoted with overlaid dots labelled "1" and "2".

**Movie S5 – Surface Traffic Clustering at High Curvature Edge** illustrates the initiation of two adjacent protrusions due to edge clustering in a region of locally higher convex edge curvature, thus demonstrating one common way in which protrusions were observed forming. The scale bar represents 1 $\ell$.

**Movie S6 – Surface Traffic on Young Protrusion** illustrates the growth of a relatively young, and highly tapered protrusion. In such a case, the raft geometry effectively funnels surface ants to the tip where they cluster and promote runaway growth. The scale bar represents 1 $\ell$.

**Movie S7 – Surface Traffic on Mature Protrusion** illustrates the growth of a mature, tether-like protrusion and visually illustrates the directional traffic occurring on this feature. The scale bar represents $1\,\ell$.

**Movie S8 – Surface Traffic and Spontaneous Protrusion Initiation** illustrates the initiation of a protrusion in a raft region with a relatively low edge curvature, indicating that while convex edge curvature may facilitate asymmetric growth, there are other factors such as cooperative motion of surface ants that can lead to breaks in symmetry. The scale bar represents $1\,\ell$.

**Movie S9 – Image Tracking Sample** illustrates a region of interest in which the trajectory of every surface ant was manually tracked to collect velocity data.